\documentclass{article}
\usepackage{graphicx}
\usepackage[english]{babel}

\usepackage{amsmath}
\usepackage{float}
\usepackage{dcolumn}
\usepackage{bm}
\usepackage{color}

\title{
\includegraphics[width=0.35\textwidth]{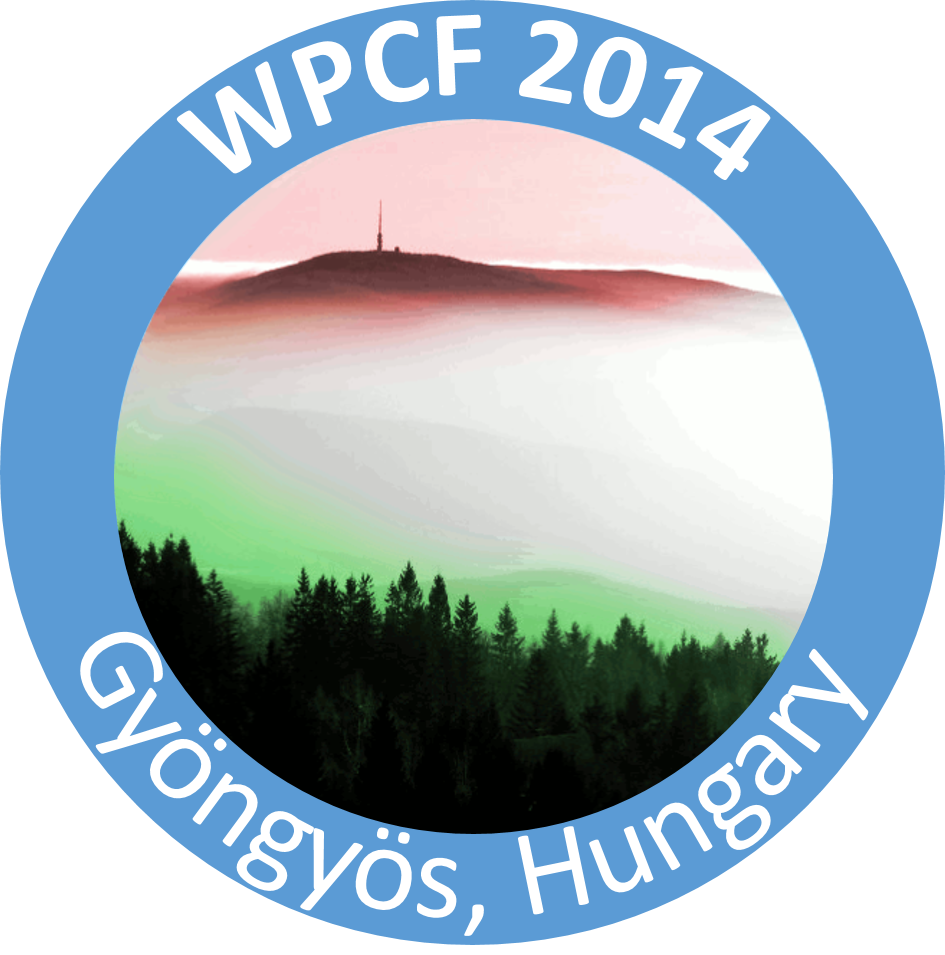}\\[1cm]
Pion Transverse Momentum Spectrum, Elliptic Flow and Interferometry in the
Granular Source Model in Ultra-Relativistic Heavy Ion Collisions}

\author{{Jing Yang$^1$, Yan-Yu Ren$^2$ and Wei-Ning Zhang$^{1,2}$}\\[1ex]
$^1$School of Physics and Optoelectronic Technology, \\Dalian University of
Technology, Dalian, China\\
$^2$Department of Physics, Harbin Institute of Technology, Harbin, China\\
}

\begin{document}

\fontfamily{lmss}\selectfont
\maketitle

\begin{abstract}
We systematically investigate the pion transverse momentum spectrum, elliptic
flow, and Hanbury-Brown-Twiss (HBT) interferometry in the granular source model
of quark-gluon plasma droplets in ultra-relativistic heavy ion collisions.
The granular source model can well reproduce the experimental results of
the Au-Au collisions at $\sqrt{s_{NN}}=$ 200 GeV and the Pb-Pb collisions
at $\sqrt{s_{NN}} =$ 2.76 TeV with different centralities.  We examine the
parameters of the granular source models with an uniform and Woods-Saxon
initial energy distributions in a droplet.  The parameters exhibit certain
regularities for collision centrality and energy.
\end{abstract}

\section{Introduction}

Single particle transverse momentum spectrum, elliptic flow, and Hanbury-Brown-Twiss
(HBT) interferometry are three important final particle observables in high energy
heavy ion collisions.  They reflect the characters of the particle-emitting sources
in different aspects and at different stages.  Therefore, a combined investigation
of these observables can provide very strong constrains for source models.

So far, much progress has been made in understanding the experimental data of the
heavy ion collisions at the top energies of the Relativistic Heavy Ion Collider (RHIC)
\cite{RHIC_exp-rep,QM06exp,QM06theor,QM08exp,QM08sum,QM09sum,QM11exp,QM11theor,QM12sum}.
However, more detailed investigations of the physics beneath the data through
multiobservable analyses are still needed.  On the other hand, the experimental data
of the Pb-Pb collisions at $\sqrt{s_{NN}}=$ 2.76 TeV at the Large Hadron Collider
(LHC) have been recently published \cite{ALI-spe13z,ALI-v2-11z,ALI-hbt11,ALI-hbt11z}.
It is an ambitious goal for models to explain the experimental data of particle spectra,
elliptic flow, and HBT interferometry in different centrality region consistently for
the heavy ion collisions at the RHIC and LHC.

In Refs. \cite{WNZ04,WNZ06,WNZ09}, W. N. Zhang {\it et al.} proposed and developed
a granular source model of quark-gluon plasma (QGP) droplets to explain the HBT data
of the RHIC experiments \cite{STA-hbt01,PHE-hbt02,PHE-hbt04,STA-hbt05z}.
In Ref. \cite{WNZ11J,WNZ11}, the granular source model was used to explain the pion
transverse momentum spectrum and HBT data of the most central heavy ion collisions
at the RHIC and LHC.  Motivated by these successes, we systematically investigate
the pion transverse momentum spectrum, elliptic flow, and HBT interferometry for
the granular sources in the heavy ion collisions at the RHIC and LHC energies with
different centralities.  The granular source parameters for an uniform and Woods-Saxon
initial energy distributions in a droplet are examined and compared.

\section{Granular Source Model}

The granular source model of QGP droplets regards the whole source evolution as
the superposition of the evolutions of many QGP droplets.  Each droplet has a
position-dependent initial velocity and evolves hydrodynamically.  The model
construction is based on the following suggestions.  In the heavy ion collisions
at top RHIC energies and LHC energies, the created strong coupled QGP (sQGP) systems
at central rapidity region may reach local equilibrium at a very short time, and
then expand rapidly along the beam direction ($z$-axis).  Because the local
equilibrium system is not uniform in the transverse plane ($x$-$y$ plane)
\cite{ISFFSC12}.  The system may form many tubes along the beam direction during
the subsequent fast longitudinal expansion and finally fragment into many QGP
droplets with the effects of ``sausage" instability, surface tension, and bulk
viscosity \cite{WNZ06,CYW73,Tor08,Tak09}.

As in Ref. \cite{WNZ06}, we suppose the QGP droplets of the granular source
initially distribute within a cylinder along $z$-axis by
\begin{eqnarray}
\frac{dN_d}{dx_0dy_0dz_0}\!\! &{\propto}&\!\!\Big[1-e^{-(x_0^2+y_0^2)/
\Delta {\cal R}_T^2}\Big]\theta({\cal R}_T- \rho_0)\cr
&&\times \theta({\cal R}_z -|z_0|).
\end{eqnarray}
Here $\rho_0=\sqrt{x_0^2+y_0^2}$ and $z_0$ are the initial transverse and
longitudinal coordinates of the droplet centers.  The parameters ${\cal R}_T$
and ${\cal R}_z$ describe the initial transverse and longitudinal sizes of the
source, and $\Delta {\cal R}_T$ is a transverse shell parameter \cite{WNZ06}.

In Ref. \cite{WNZ11}, the Bjorken hypothesis \cite{Bjo83} of longitudinal
boost-invariant is used to describe the longitudinal velocity of droplet for
the most central collisions, and the transverse velocity of droplet has a form
of exponential power.  Considering the longitudinal velocity of droplet varying
with collision centrality, we introduce here also a longitudinal power parameter,
which will be determined by experimental data, to describe the longitudinal
velocity phenomenologically.  The initial velocities of the droplets in granular
source frame are assumed as \cite{WNZ06}
\begin{equation}
\label{ini_v}
v_{{\rm d}i}=\mathrm{sign}(r_{0i}) \cdot a_i \bigg(\frac{|r_{0i}|}{{\cal R}_i}
\bigg)^{b_i},~~~~~~i=1,\,2,\,3,
\end{equation}
where $r_{0i}$ is $x_0$, $y_0$, or $z_0$ for $i=$ 1, 2, or 3, and $\text{sign}
(r_{0i})$ denotes the signal of $r_{0i}$, which ensures an outward droplet
velocity.  In Eq. (\ref{ini_v}), ${\cal R}_i=({\cal R}_T,{\cal R}_T,{\cal R}_z)$,
$a_i=(a_x,a_y,a_z)$ and $b_i=(b_x,b_y,b_z)$ are the magnitude and exponent
parameters in $x$, $y$, and $z$ directions, which are associated with the early
thermalization and pressure gradients of the system at the breakup time.  It is
also convenient to use the equivalent parameters $\overline{a}_T =(a_x+a_y)/2$
and $\Delta a_T=a_x-a_y$ instead of $a_x$ and $a_y$.  The parameters $\overline{a}_T$
and $\Delta a_T$ describe the transverse expansion and asymmetric dynamical behavior
of the system at the breakup time, respectively.  For simplicity, we take $b_x=b_y=b_T$
in calculations.  The parameters $b_T$ and $b_z$ describe the coordinate dependence of
exponential power in transverse and longitudinal directions.

In the calculations of the hydrodynamical evolution of the droplet, we use the
equation of state (EOS) of the S95p-PCE165-v0 \cite{She10}, which combines the
lattice QCD data at high temperature with the hadron resonance gas at low
temperature.  We assume systems fragment when reaching a certain local energy
density $\epsilon_0$, and take the initial energy density of the droplets to
be 2.2 GeV/fm$^3$ for all considered collisions for simplicity \cite{WNZ11}.
The initial droplet radius $r_0$ is supposed satisfying a Gaussian distribution
with the standard deviation $\sigma_d=$ 2.5 fm in the droplet local frame
\cite{WNZ11}.  We consider an uniform initial energy distribution in droplet
and a Woods-Saxon distribution,
\begin{equation}
\epsilon(r)=\epsilon_0\frac{1}{e^{\frac{r-r_0}{a}}+1}\,,
\label{Eq-woodssaxon}
\end{equation}
where $a=0.1r_0$.

With the evolution of the hot droplets, the final pions freeze out at temperature
$T_f$ with the momenta obeying Bose-Einstein distribution.  To include the resonance
decayed pions emitted later as well as the directly produced pions at chemical
freeze out earlier, a wide region of $T_f$ is considered with the probability
\cite{WNZ11}
\begin{eqnarray}
\frac{dP}{dT_f} &{\propto}& f_{\text{dir}}\,e^{-\frac{T_{\text{chem}}-T_f}
{\Delta T_{\text{dir}}}}+({1-f_{\text{dir}}})\cr
&&\times e^{-\frac{T_{\text{chem}}-T_f}{\Delta T_{\text{dec}}}},
~~(T_{\text{chem}}>T_f>80~\text{MeV}),
\end{eqnarray}
where $f_{\text{dir}}$ is the fraction of the direct emission around the chemical
freeze out temperature $T_{\text{chem}}$,  $\Delta T_{\text{dir}}$ and $\Delta
T_{\text{dec}}$ are the temperature widths for the direct and decay emissions,
respectively.  In the calculations, we take $f_{\text{dir}}=0.75$, $\Delta
T_{\text{dir}}=10$ MeV, and $\Delta T_{\text{dec}}=90$ MeV as in Ref. \cite{WNZ11}.
The value of $T_{\text{chem}}$ is taken to be 165 MeV as it be taken in the
S95p-PCE165-v0 EOS \cite{She10}.

After fixing the parameters used in the calculations of hydrodynamical evolution
and freeze-out temperature, the free model parameters are the three source geometry
parameters ($R_T$, $\Delta R_T$, $R_z$) and the five droplet velocity parameters
($\overline{a}_T$, $\Delta a_T$, $a_z$, $b_T$, $b_z$).  They are associated with
the initial sizes, expansion, and directional asymmetry of system, and have
significant influence on the observables of pion momentum sprectra, elliptic flow,
and HBT radii in the granular source model.  In next section, we will determine
these parameters by the experimental data of these observable, and examine their
variations with collision energy and centrality for the heavy ion collisions at
the RHIC and LHC.

\section{Results of Pion Momentum Spectrum, Elliptic Flow and Interferometry}

\begin{figure}[!htb]
\begin{center}
\includegraphics[angle=0,scale=0.55]{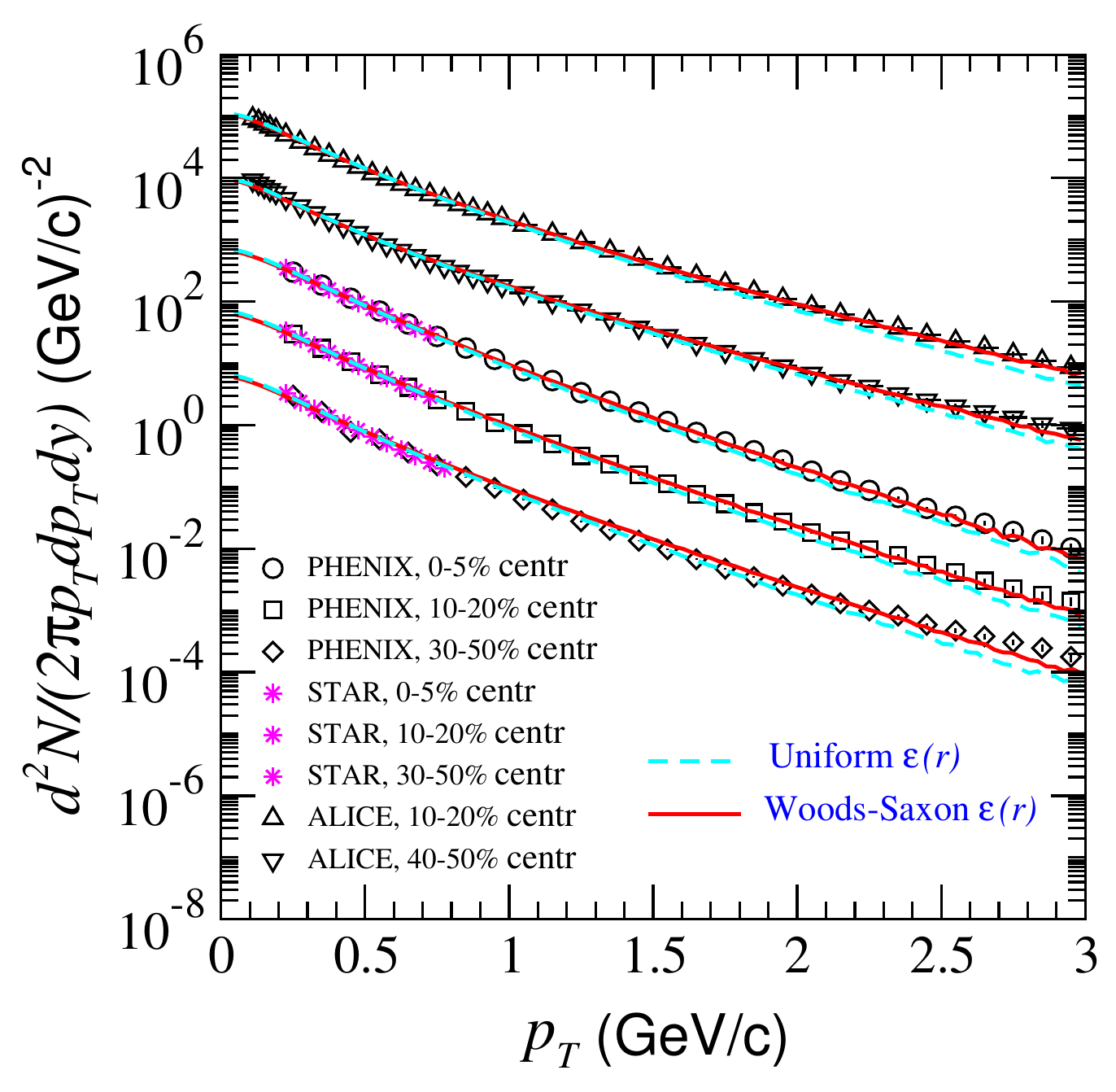}
\caption{(Color online) The pion transverse momentum spectra of the granular
sources for the RHIC Au-Au collisions at $\sqrt{s_{NN}}=$ 200 GeV and the LHC
Pb-Pb collisions at $\sqrt{s_{NN}}=$ 2.76 TeV, for the uniform and Woods-Saxon
initial energy distributions in a droplet.  The experimental data of PHENIX
\cite{PHE-spe04z}, STAR \cite{STA-spe04z}, and ALICE \cite{ALI-spe13z} are
also plotted. }
\label{zfsp}
\end{center}
\end{figure}

In high energy heavy ion collisions, the invariant momentum distribution of final
particles can be written in the form of a Fourier series \cite{SV-YZ96,AP-SV98},
\begin{eqnarray}
\label{pdis}
E\frac{d^3N}{d^3p}=\frac{1}{2\pi}\frac{d^2N}{p_Tdp_Tdy}\left[1+\sum_n{2v_n
\cos(n\phi)}\right],
\end{eqnarray}
where $E$ is the energy of the particle, $p_T$ is the transverse momentum, $y$ is
the rapidity, and $\phi$ is the azimuthal angle with respect to the reaction plane.
In Eq. (\ref{pdis}), the first term on right is the transverse momentum spectrum
in the rapidity region $dy$, and the second harmonic coefficient $v_2$ in the
summation is called elliptic flow.

\begin{figure}[!htb]
\begin{center}
\vspace*{5mm}
\includegraphics[angle=0,scale=0.5]{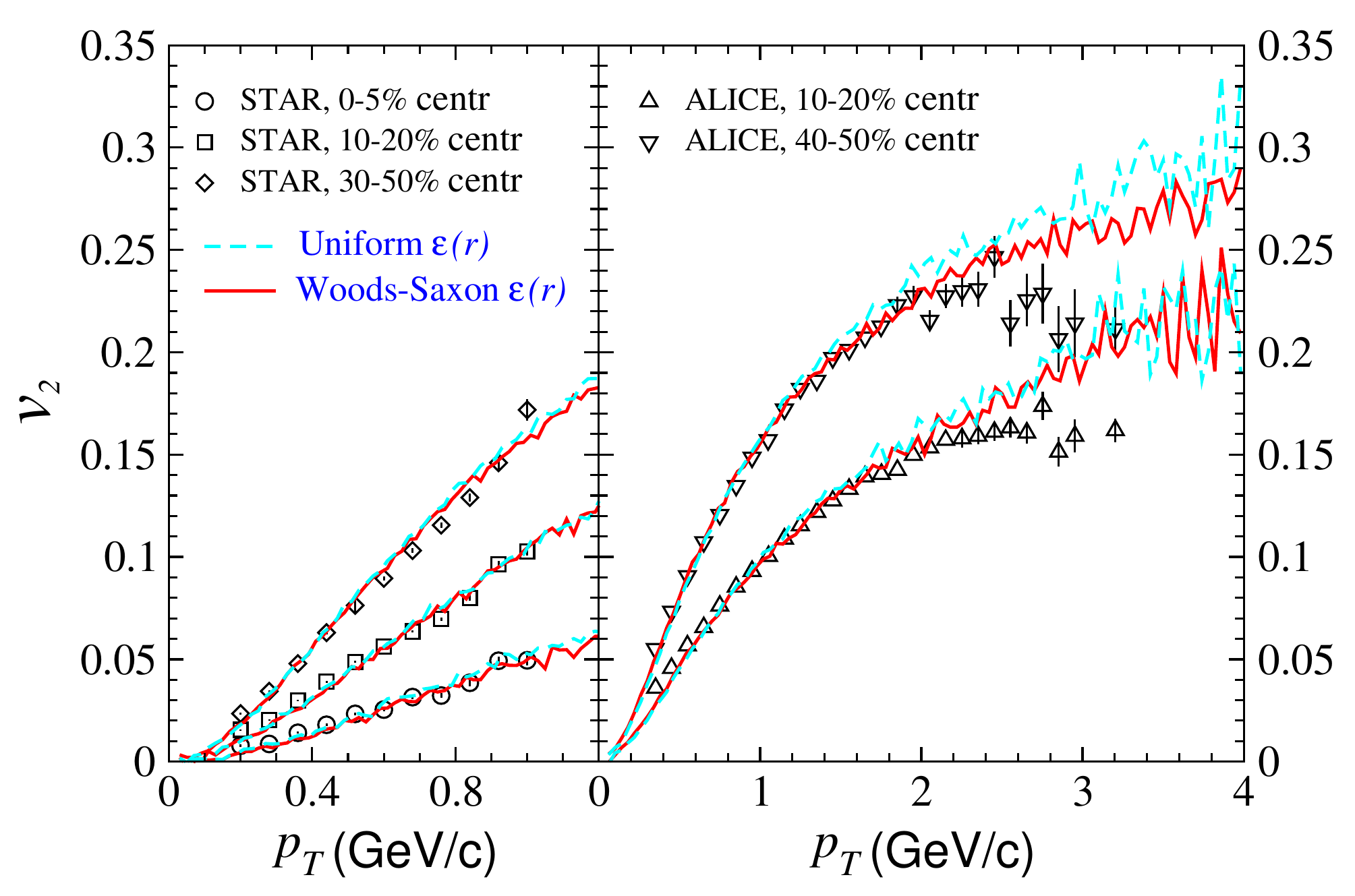}
\caption{(Color online) The pion elliptic flow of the granular sources for the
RHIC Au-Au collisions at $\sqrt{s_{NN}}=$ 200 GeV and the LHC Pb-Pb collisions
at $\sqrt{s_{NN}}=$ 2.76 TeV, for the uniform and Woods-Saxon initial energy
distributions in a droplet.  The experimental data of STAR \cite{STA-v2-05z},
and ALICE \cite{ALI-v2-11z} are also plotted. }
\label{zfv2}
\end{center}
\end{figure}

In Fig. \ref{zfsp}, we plot the pion transverse momentum spectra of the granular
sources with the uniform and Woods-Saxon initial energy distributions in a droplet.
The experimental data of the Au-Au collisions at $\sqrt{s_{NN}}=$ 200 GeV at the
RHIC \cite{PHE-spe04z,STA-spe04z} and the Pb-Pb collisions at $\sqrt{s_{NN}}=$
2.76 TeV at the LHC \cite{ALI-spe13z} are also plotted.
In Fig. \ref{zfv2}, we plot the pion elliptic flow results of the granular sources
with the uniform and Woods-Saxon initial energy distributions, and the experimental
data of the Au-Au collisions \cite{STA-v2-05z} and the Pb-Pb collisions
\cite{ALI-v2-11z}.

The transverse momentum spectrum and elliptic flow of the granular sources
are well in agreement with the experimental data, except for the elliptic
flow results at $p_T>2$ GeV/$c$.  The differences between the results of
the granular sources with the uniform and Woods-Saxon initial energy
distributions are small.  The experimental data of the momentum spectrum
and elliptic flow at the same centralities can simultaneously give the
strong constraints to the velocity parameters of the granular sources.
After then, the geometry parameters of the granular sources for the
collisions with the different centralities can be further determined by
the experimental data of HBT interferometry at the same centralities.

\begin{figure}[!htbp]
\begin{center}
\vspace*{5mm}
\includegraphics[scale=0.47]{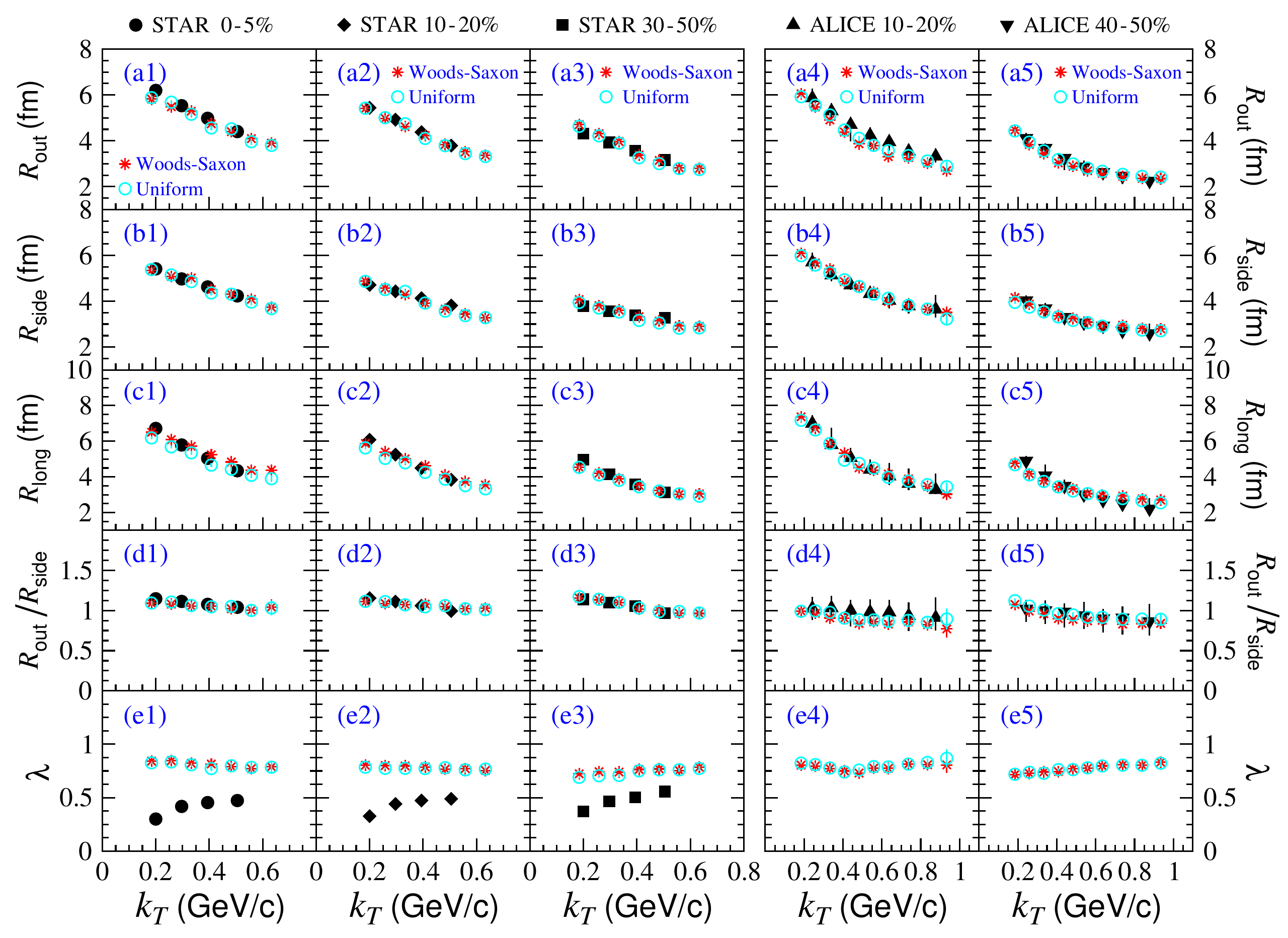}
\caption{(Color online) The HBT results of the granular sources for the uniform
and Woods-Saxon initial energy distribution, and the experimental data for the
Au-Au collisions at the RHIC \cite{STA-hbt05z} and the Pb-Pb collisions at the
LHC \cite{ALI-hbt11z} with different centralities. }
\label{zfhbtfit}
\end{center}
\end{figure}

Two-particle HBT correlation function is defined as the ratio of the
two-particle momentum spectrum $P({\bf p}_1,{\bf p}_2)$ to the product
of two single-particle momentum spectra $P({\bf p}_1)P({\bf p}_2)$.
It has been widely used to extract the space-time geometry, dynamic
and coherence information of the particle-emitting source in high
energy heavy ion collisions \cite{Gyu79,Wongbook,Wie99,Wei00,Lisa05}.
In the usual HBT analysis in high energy heavy ion collisions, the
two-pion correlation functions are fitted by the Gaussian parameterized
formula
\begin{equation}
\label{CF}
C(q_{\rm out},q_{\rm side},q_{\rm long})\!=\!1 \!+ \lambda\,
e^{-R_{\rm out}^2 q_{\rm out}^2 -R_{\rm side}^2 q_{\rm side}^2
-R_{\rm long}^2 q_{\rm long}^2},
\end{equation}
where $q_{\rm out}$, $q_{\rm side}$, and $q_{\rm long}$ are the Bertsch-Pratt
variables \cite{Ber88,Pra90}, which denote the components of the relative
momentum ${\bf q}={\bf p}_1-{\bf p}_2$ in transverse out and side directions
and in longitudinal direction, respectively.  In Eq. (\ref{CF}) $\lambda$ is
chaoticity parameter of source, $R_{\rm out}$, $R_{\rm side}$, and $R_{\rm long}$
are the HBT radii in out, side, and long directions.

We plot in Fig. \ref{zfhbtfit} the two-pion HBT results for the granular sources
and the experimental data of the RHIC Au-Au collisions \cite{STA-hbt05z} and the
LHC Pb-Pb collisions \cite{ALI-hbt11z} with the same centralities as the
experimental data of the spectrum and elliptic flow.  Here, $k_T$ is the transverse
momentum of pion pair.  One can see that the granular source models can well reproduce
the experimental HBT radii and their variations with $k_T$.
The results of the chaotic parameters $\lambda$ of the granular sources are larger
than the experimental data, because there are many effects in experiments can
decrease $\lambda$ \cite{Gyu79,Wongbook,Wie99,Wei00,Lisa05}, which exceed our
considerations.

In Table 1, we present the parameters of the granular sources with uniform
initial energy distribution in a droplet, determined by the experimental data
of the momentum spectra, elliptic flow, and HBT radii.  The values of the
source geometry parameters indicate that the sources are a short cylinder with
small shell effect (small $\Delta R_T$).  For certain collision energy, the
source geometry parameters $R_T$, $\Delta R_T$ and $R_z$ increase with the
collision centralities.  However, for the 10-20\% centrality, these geometry
parameters for the LHC collisions are larger than those for the RHIC collisions.
The large difference between the droplet velocity parameters $b_T$ and $b_z$
indicates the different dynamical behaviors of the sources in the transverse
and long directions.  The parameter $\Delta a_T$ increases with decreasing
centrality. And, the values of $\overline{a}_T$ and $a_z$ are almost independent
of collision centrality.  A detail analysis on the relationships between the
source parameters and the granular source space-time evolution can be seen
in Ref. \cite{JYANG14}.

\begin{table}[!hbt]
\begin{center}
\caption{The geometry parameters (in fm unit) and velocity parameters of the granular
sources with uniform initial energy distribution in a droplet. }
\begin{tabular}{c|ccccccccc}
\hline\hline
Centrality&~${\cal R}_T$&$\Delta {\cal R}_T$&${\cal R}_z$&
$\overline{a}_T$&$\Delta a_T$&$a_z$&$b_T$&$b_z$\\
\hline
RHIC,~~~0--5~\% & 5.8 & 0.7 & 3.9 & ~0.469 & ~0.066 & ~0.593 & ~0.76 & ~0.13 \\
RHIC,~10--20\%  & 4.5 & 0.5 & 2.9 & ~0.454 & ~0.115 & ~0.593 & ~0.56 & ~0.11 \\
RHIC,~30--50\%  & 2.5 & 0.3 & 0.5 & ~0.437 & ~0.156 & ~0.593 & ~0.37 & ~0.06 \\
\hline
\,LHC,\,~10--20\% & 6.0 & 0.9 & 5.5 & ~0.431 & ~0.092 & ~0.592 & ~0.35 & ~0.13 \\
\,LHC,\,~40--50\% & 2.5 & 0.4 & 1.8 & ~0.407 & ~0.131 & ~0.590 & ~0.23 & ~0.03 \\
\hline\hline
\end{tabular}
\end{center}
\end{table}

\begin{table}[!htb]
\begin{center}
\caption{The geometry parameters (in fm unit) and velocity parameters of the granular
sources with Woods-Saxon initial energy distribution in a droplet. }
\vspace*{2mm}
\begin{tabular}{c|ccccccccc}
\hline\hline
Centrality&~${\cal R}_T$&$\Delta {\cal R}_T$&${\cal R}_z$&
$\overline{a}_T$&$\Delta a_T$&$a_z$&$b_T$&$b_z$\\
\hline
RHIC,~~~0--5~\% & 5.8 & 0.7 & 5.1 & ~0.469 & ~0.066 & ~0.52 & ~0.76 & ~0.13 \\
RHIC,~10--20\%  & 4.5 & 0.5 & 4.0 & ~0.457 & ~0.122 & ~0.52 & ~0.56 & ~0.11 \\
RHIC,~30--50\%  & 2.8 & 0.3 & 1.8 & ~0.453 & ~0.156 & ~0.52 & ~0.37 & ~0.06 \\
\hline
\,LHC,\,~10--20\% & 6.0 & 0.9 & 5.5 & ~0.496 & ~0.092 & ~0.59 & ~0.43 & ~0.13 \\
\,LHC,\,~40--50\% & 2.5 & 0.4 & 1.8 & ~0.434 & ~0.127 & ~0.59 & ~0.23 & ~0.03 \\
\hline\hline
\end{tabular}
\end{center}
\end{table}

In Table 2, we present the parameters of the granular sources with Woods-Saxon
distribution of initial energy in a droplet, determined by the experimental data
of the momentum spectra, elliptic flow, and HBT radii.  By Comparing the two set
parameters in Table 1 and Table 2, one can see that there are some differences
between the $R_z$ and $a_z$ values for the RHIC collisions.  Also, for the LHC
collisions, the values of $\bar{a}_T$ for for the granular sources with the
Woods-Saxon distribution are larger.

\section{Summary and Conclusions}

We systemically investigate the pion transverse momentum spectrum, elliptic flow,
and HBT interferometry in the granular source model for the heavy ion collisions
at the RHIC highest energy and the LHC energy.  The centrality dependence of the
observables at the two energies are examined.  By comparing the granular source
results with the experimental data of the Au-Au collisions at $\sqrt{s_{NN}}=200$
GeV at the RHIC and the Pb-Pb collisions at $\sqrt{s_{NN}}=2.76$ TeV at the LHC
with different collision centralities, we investigate the geometry and velocity
parameters in the granular source models with an uniform and Woods-Saxon initial
energy distributions in a droplet.  The parameters as a function of collision
centrality and energy are examined.  Our investigations indicate that the granular
source model can well reproduce the experimental data of pion transverse momentum
spectra, elliptic flow, and HBT radii in the Au-Au collisions at $\sqrt{s_{NN}}=$
200 GeV with 0--5\%, 10--20\%, and 30--50\% centralities
\cite{STA-hbt05z,PHE-spe04z,STA-spe04z,STA-v2-05z}, and in the Pb-Pb collisions
at $\sqrt{s_{NN}}=$ 2.76 TeV with 10--20\% and 40--50\% centralities
\cite{ALI-spe13z,ALI-v2-11z,ALI-hbt11z}.  The experimental data of pion momentum
spectra, elliptic flow, and HBT radii impose very strict constraints on the
parameters in the granular source model.  They exhibit certain regularities for
collision centrality and energy.

\vspace*{0.5cm}
\noindent{\Large \bf Acknowledgement}

\vspace*{0.3cm} This work was supported by the National Natural Science Foundation of
China, Contract No. 11275037.


\end{document}